\documentclass[journal]{IEEEtran}
\usepackage{graphicx}
\usepackage{multirow}
\usepackage{subfigure}

\usepackage[belowskip=-15pt,aboveskip=0pt]{caption}

\ifCLASSINFOpdf

\else

\fi

\begin{document}
%
\title{Evaluating Impact of Mobility \\
on Wireless Routing Protocols}

\author{\IEEEauthorblockN{N. Javaid$^{\dag,\$}$, M. Yousaf$^{\$}$, A. Ahmad$^{\ddag}$, A. Naveed$^{\ddag}$, K.     Djouani$^{\dag,\S}$\\\vspace{0.4cm}}
    \IEEEauthorblockA{ $^{\dag}$LISSI, Universit\'e Paris-Est Cr\'eteil (UPEC), France. \{nadeem.javaid,djouani@univ-paris12.fr\}}\\
    $^{\$}$Dept. of Electrical Engineering, COMSATS, Islamabad, Pakistan. \{nadeemjavaid@comsats.edu.pk\}\\
    $^{\ddag}$Institute of Computing and Information Technology, Gomal University, D.I.Khan, Pakistan. \\
    $^{\S}$French South African Institute of Technology , Pretoria, South Africa. \{djouanik@tut.ac.za\}
     }

\maketitle

\begin{abstract}
In this paper, we evaluate, analyze, and compare the impact of mobility on the behavior of three reactive protocols (AODV, DSR, DYMO) and three proactive protocols (DSDV, FSR, OLSR) in multi-hop wireless networks. We take into account throughput, end-to-end delay, and normalized routing load as performance parameters. Based upon the extensive simulation results in NS-2, we rank all of six protocols according to the performance parameters. Besides providing the interesting facts regarding the response of each protocol on varying mobilities and speeds, we also study the trade-offs, the routing protocols have to make. Such as, to achieve throughput, a protocol has to pay some cost in the form of increased end-to-end delay or routing overhead.
\end{abstract}

\begin{IEEEkeywords}
AODV, DSDV, DSR, DYMO, FSR, OLSR, throughput, end-to-end delay, normalized routing load, wireless multi-hop networks, mobility
\end{IEEEkeywords}

\IEEEpeerreviewmaketitle
\vspace{-0.4cm}
\section{Introduction}


	To correctly illustrate the performance evaluation of the routing protocols, it is remarkably significant to exactly depict the movement of mobile nodes. So, Shams  \textit{et al.} in \cite{3-1} designed scenario-based mobility models which closely present the movement patterns of users in real life and they have evaluated two reactive routing protocols, AODV and DSDV. The proposed mobility models are: Fast Car Model (FCM), Slow Car Model (SCM), Human Running Model (HRM) and Human Walking Model (HWM). We follow the same models for this study. FCM states that the mobile nodes are vehicles moving up to the speeds of 30m/s or 108km/h \cite{3-2} on highways and motor ways. In practice, vehicles do not move with this speed all the time rather they take pauses at different break points and traffic signals. Thus 'pause-time' intervals are also considered. Like FCM, SCM also considers the vehicles but moving with the speed of 15m/s or 45km/h on the busy roads and cannot move at higher speeds. It is observed that most of the times, wireless devices are carried by the humans. For example, soldiers in the combat zone can run or walk, people jogging on different tracks, in emergency situations, sports and so on. In short, 8m/s or 28.8km/h can be taken as an average speed for SCM. The HWM is identical to the HRM model but with an average speed of 2m/s or 7.2km/h \cite{3-2}. The examples for HWM may be people walking in the shopping centers, university or college campuses, etc.

	Being an interface between the underlying wireless network and mobile users, a routing protocol plays an important role. So, to provide the reader with a comprehensive idea about routing and how do the routing protocols react to the topological changes, we have chosen the most widely experimented and frequently used protocols for our study; three from reactive or on-demand class: Ad-hoc On-demand Distance Vector (AODV), Dynamic Source Routing (DSR), DYnamic MANET On-demand (DYMO), and three from proactive or table-driven class Destination Sequenced Distance Vector (DSDV), Fish-eye State Routing (FSR), Optimized Link State Routing (OLSR). Authors in \cite{3-1}, have analysed two protocols; AODV and DSDV. Simulations are run for four pause times (0s, 1s, 10s and 450s). However, routing protocols being categorized in reactive and proactive classes are yet to be analyzed. Moreover, to perform a precise and detailed analysis we have simulated six protocols with ten pause times (0s, 100s, 200s,. . . and 900s).

\vspace{-0.4cm}
\section{Routing protocols and mobility}
This section is devoted to short description of each protocol, stating the routing technique working behind it, class to which the protocol belongs; i. e., reactive or proactive, the way in which it performs \textit{route discovery (RD)}, \textit{route maintenance (RM)}, \textit{route table (RT) calculation} and at the end, the claims made by each protocol to deal with the mobility. At the end of section, Table.1 summarizes all of the six protocols.
\vspace{-0.4cm}
\subsection{Reactive protocols and mobility}
AODV \cite{3-3}, 
\cite{3-4}, 
DSR \cite{3-3}, 
\cite{3-5}, 
\cite{3-6}  
and DYMO \cite{3-7}, 
\cite{3-8} 
are multi-hop on-demand routing protocols. Their on-demand nature has a great impact on mobility because they compute routes only when needed making them suitable for mobile scenarios. AODV claims that "it can handle low, moderate, and relatively high mobility rates, as well as a variety of data traffic levels" \cite{3-4}. DSR claims that "it adapts quickly to the topological changes when movement of nodes is frequent. It requires little or no routing overhead during the periods in which nodes move less frequently or remain at rest" \cite{3-6}. DYMO states that ``it adapts to changing network topology and determines unicast routes between nodes within the network in 'on-demand' fashion" \cite{3-7}. So, in this study we evaluate and compare the performance of these protocols based upon their claims regarding mobility. All of three protocols use flooding based \textit{RD}for path calculation, as shown in Fig.1. AODV uses hop-by-hop routing while DSR and DYMO use source routing as packet forwarding scheme. These protocols implement two common operations: \textit{RD} and \textit{RM}.

\begin{figure}[!h]
\centering
\includegraphics[
height=18 cm,
width=8 cm
]{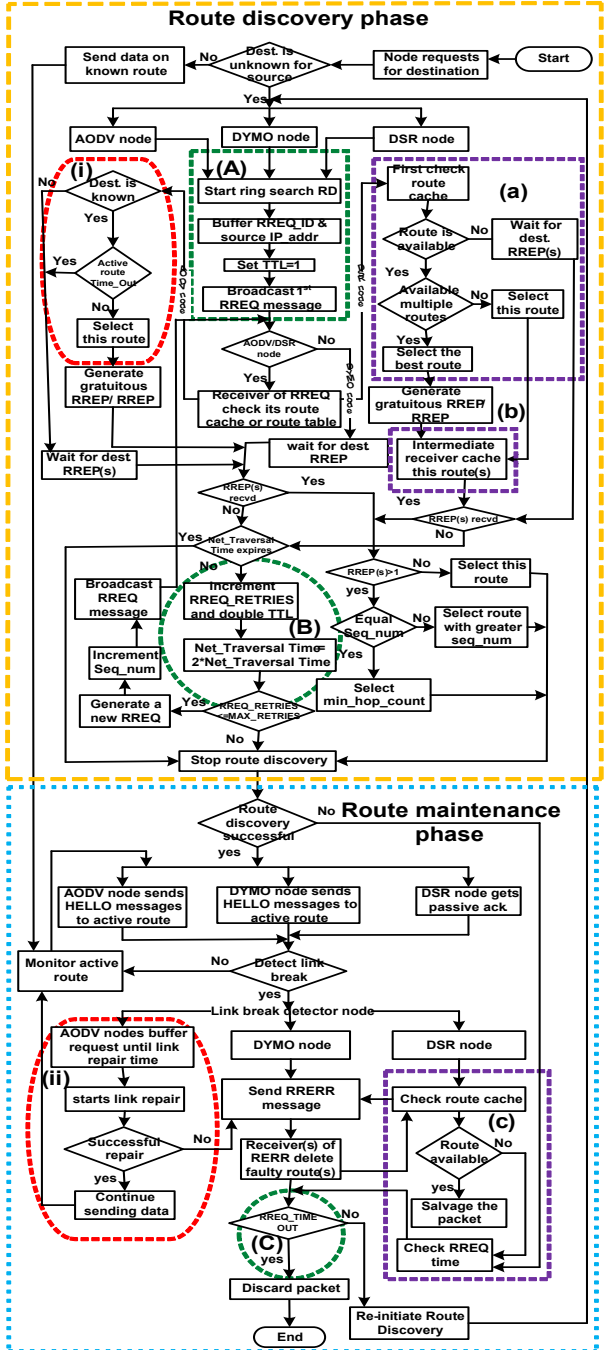}
\caption{Reactive protocols, 'route discovery' and 'route maintenance'.}
\end{figure}

\subsection{Proactive Protocols and Mobility}

DSDV \cite{3-9}, 
FSR \cite{3-10}, 
\cite{3-11}       
and OLSR \cite{3-12}, 
\cite{3-13}, 
are table-driven proactive protocols. All of these proactive protocols use \textit{hop-by-hop routing} scheme for packet forwarding. In DSDV, distance vector packets are dispersed and then \textit{Distributed Bellman Ford (DBF)} algorithm is used for path calculation, as shown in Fig.2. In FSR, DBF algorithm is used for path calculation and link state packets are not flooded. The nodes maintain a link state table based on up-to-date information received from the neighboring nodes and they periodically exchange it with their local neighbors only.

\begin{figure}[!h]
\centering
\includegraphics[
height=16 cm,
width=8 cm
]{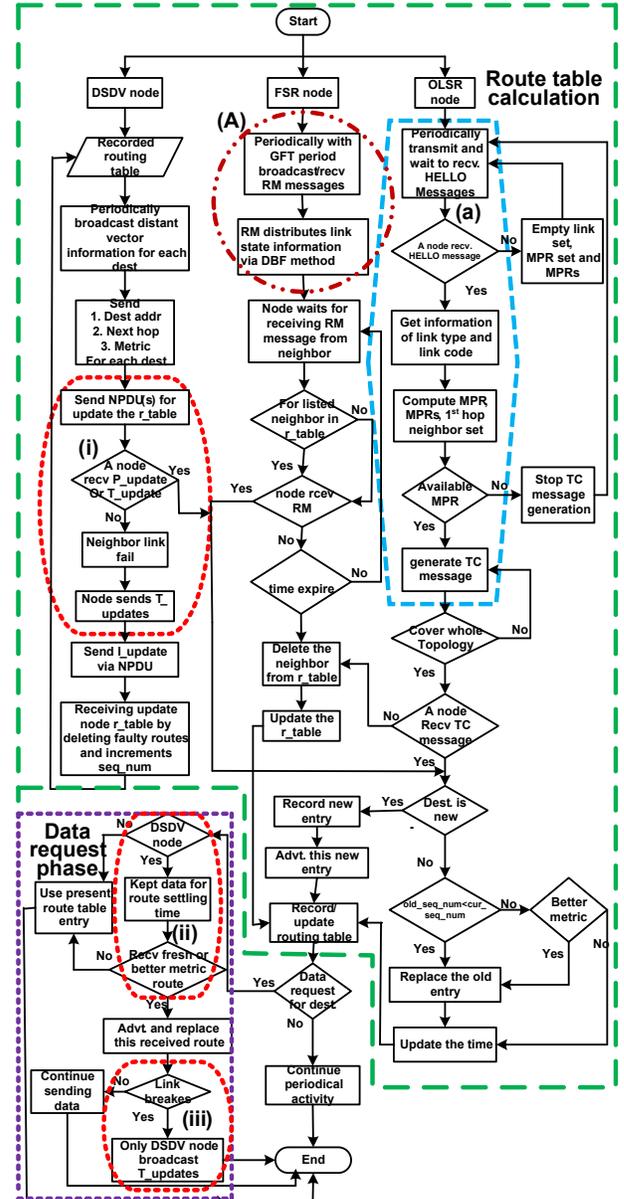}
\caption{Proactive protocols, 'route calculation' and 'data request'}%
\label{fig1}%
\end{figure}

For the path calculation OLSR uses \textit{Dijkstra's algorithm}. To maintain consistency in routing tables, DSDV generates periodic updates, $P_updates$ and trigger updates, $T_updates$, when information about new links becomes available. For convergence, routing information is advertised by broadcasting the packets periodically.
FSR uses \textit{graded-frequency (GF)} mechanism to achieve route accuracy while \textit{Multi-point Relay (MPR)} redundancy mechanism is used by OLSR in high dynamic situations. DSDV uses two types of packets: first type carries all available routing information, called a \textit{full dump} and second type carries only information changed since the last full dump, called an \textit{incremental}. An \textit{incremental} should fit in an \textit{Network Protocol Data Units (NPDUs)}, as illustrated in step(i) of Fig.2. Moreover, \textit{(NPDUs)} are used to control the network overhead, by arranging the "incremental" and "full dumps" utilizing the bandwidth.


\begin{table}[h]
\caption {Routing Protocols in brief}
\centering
\tiny\begin{tabular}{|c|c|c|c|c|c|}
\hline
\multirow{2}{*}{\textbf{Protocol}}&\textbf{Distnguishd}&\textbf{Path}&\textbf{Packet}&
\textbf{Flooding cntrl}&\textbf{Overhead}\\
        &\textbf{features}&\textbf{calculation}&\textbf{forwarding}&
        \textbf{mechanism}&\textbf{reduction}\\
\hline

&Local link&Flooding-based&Hop-by-hop&Ring search&Exp. back-off\\
	\textbf{AODV}&repair&route discovery&Routing&algorithm&alg. and\\
    & & & & &grat. RREPs\\
\hline

&T-updates  &DBF      &Hop-by-hop&Exchng toplgy&Incremental\\
   \textbf{DSDV} &along with       &algorithm&routing   &info. with       &updates\\
    &P-updates &         &          &nghbrs only   &  \\

 \hline

	&Pckt salvaging &Flooding-based &	Source &Ring search &Exp.back-off \\
\textbf{DSR}&	of route &route &Routing&algorithm&alg.and\\
&cache&discovery&&&pckt salvagng\\
 \hline

\multirow{2}{*}{\textbf{DYMO}}&	No use of&Flooding-based &Source &Ring search &Exp. back-off\\
&grat. RREPs&route discovery&routing&algorithm&algorithm\\
 \hline

 \multirow{2}{*}{\textbf{FSR}}&Multi-scope&DBF&Hop-by-hop &Grdd frquncy &	 Fish-eye\\
&routing&algorithm&routing&strategy	&technique\\
 \hline

 &&Dijkstra's&Hop-by-hop &	Broadcast &\\
 \textbf{OLSR}	&MPRs&	algorithm	&routing&only through &	MPRs\\
 &&&&selected MPRs&\\

 \hline
\end{tabular}
\end{table}
\vspace{-0.5cm}

\section{Simulation Model}
	
For the simulation setup, we have chosen Continuous Bit Rate (CBR) traffic sources with a packet size of 512 bytes. The 20 source-destination pairs are spread randomly in the network. The mobility model used is Random Waypoint. The area specified is 1000m x 1000m field presenting a square space to allow the 50 mobile nodes to move inside. A square area does not "discriminate" one direction of motion like a rectangular area does. On the other hand, it limits the number of hops. (4 to 6 for a default transmission range of 250m). All of the nodes are provided with wireless links of a bandwidth of 2Mbps to transmit on. Each packet in the communication during the simulation starts its journey from a random location and moves towards a random destination with the chosen speed of 2m/s in HWM, 8m/s in HRM, 15m/s in SCM and 30m/s in FCM, as discussed in section I. Once the destination is reached, another random destination is targeted after a specified pause time (from 0s to 900s). Simulations are run for 900 seconds each.

\vspace{-0.4cm}
\section{Throughput}

It is amount of data successfully transferred from source to destination.
\vspace{-0.5cm}
\subsection{Throughput achieved by reactive protocols}

In high mobilities, DSR possess maximum throughput except in FCM at 0,100, 200 pause times, where AODV attains more throughput. In very high dynamic situations \textit{RC} of DSR becomes ineffective.

\begin{figure}[h]
\centering
\includegraphics[
height=6 cm,
width=8.5 cm
]{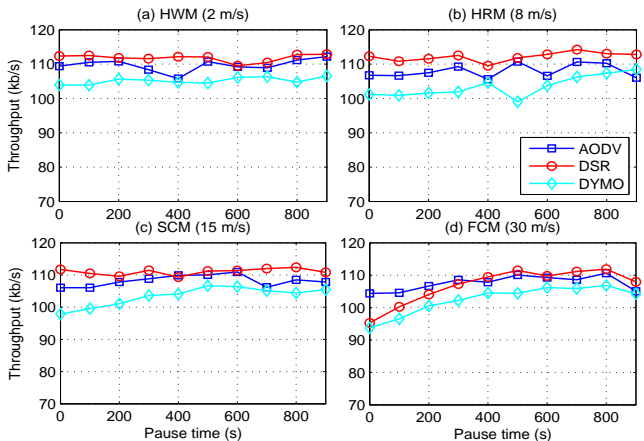}
\caption{Throughput achieved by reactive protocols for varying speeds and mobilities}
\end{figure}

As there is no mechanism to delete the \textit{stale routes} from \textit{RC} except the RERR messages; so, the protocol fails to converge at this mobility/speed. While AODV checks the \textit{route table (RT)} with valid time and avoids to use the invalid routes from routing table. The HELLO messages and \textit{LLR} make able the protocol to handle the highest rates of mobility. The overall convergence in all other situations, DSR produces the highest throughput because it does not generate more routing packets, like AODV. \textit{RC} stores multiple routes for the same destination and thus during frequent link breakage, more routes are available. Whereas, AODV's \textit{(RT)} stores one route for one destination which is also associated with a time period. Furthermore, promiscuous listening mode provides efficient mechanism to handle dynamic situation.
The worst behavior of DYMO among reactive protocols in response to mobility by showing overall less throughput value is noticed in Fig.3. The absence of \textit{grat. RREPs} and dissemination of source route information collectively result in low throughputs as compared to rest of two protocols.	

\vspace{-0.5cm}
\subsection{Throughput achieved by proactive protocols}	

Among proactive protocols, DSDV attains the highest throughput and shows efficient behavior in all mobility scenarios. The reasons for this good throughput include: firstly, when the first data packet arrives, it is kept until the best route is found for a particular destination. Secondly, a decision may delay to advertise the routes which are about to change soon, thus damping fluctuations of the route tables. The re-broadcasts of the routes with the same sequence number are minimized by delaying the advertisement of unstabilized routes. This enhances the accuracy of valid routes resulting in the increased throughput of DSDV in all types of mobility rates, as depicted in Fig.4.

Whereas, due to low convergence of OLSR in high mobility, there is a gradual decrease in overall throughput because increasing mobility increases the unavailability of valid routes due to its proactive nature. In static situation, in all of the four models, throughput is better as compared to moderate and relatively high mobility due to availability of stable entries for \textit{MPRs}. Moreover, FSR and OLSR do not trigger any control messages unlike DSDV, when links break.

\begin{figure}[h]
\centering
\includegraphics[
height=6 cm,
width=8.5 cm
]{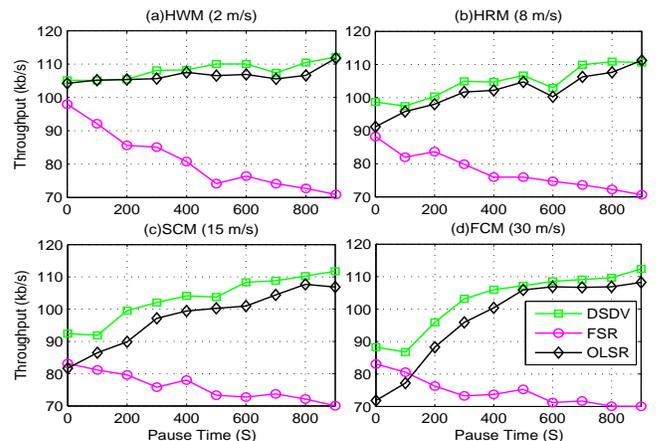}
\caption{Throughput achieved by proactive protocols for four mobility models}
\end{figure}

\vspace{-0.5cm}
\subsection{Interesting facts regarding throughput}

\textit{Reactive protocols attain more throughput than proactive ones} in high rates of mobility and speed. Reason is obvious, as proactive protocols perform route calculation before data transmission unlike the reactive ones. So, in this case if a data packet is on a calculated route and due to mobility, a link breaks, the respective proactive protocol has to perform route calculation from scratch as shown in Fig.2 that \textit{\textit{RT calculation phase}} take place first and then response to \textit{data request phase} is given, which degrades the performance. All of the six protocols achieve the throughput in the order as follows: $DSR>AODV>DSDV>DYMO>OLSR>FSR$.

\textit{DSDV sends more number of data packets than rest of the protocols} with the lowest speed of 2m/s, at 0s pause time. Because, routes with the same sequence number are not retransmitted until the route becomes stabilized, as shown in Fig.2, step(ii) in the \textit{\textit{data request phase}}.

\textit{DSDV's throughput decreases at high mobility when speed increases}. As, simultaneously increasing speed and mobility increases inconsistency in \textit{RT} calculation which leads to decrease in throughput as obvious from b, c, d, in Fig.4. DSDV achieves the same throughput values at all speeds and at moderate and no mobilities because in less mobility size of an \textit{incremental} becomes equal to size of a \textit{NPDU} to make the next \textit{incremental} smaller. For example, when a stabilized route shows a new sequence number for the same destination but the metric remains the same then this change is supposed to be non-significant and is decided to be advertised after stabilization.

\begin{figure}[h]
  \centering
 \subfigure[Scalability analysis]{\includegraphics[height=3 cm,width=4 cm]{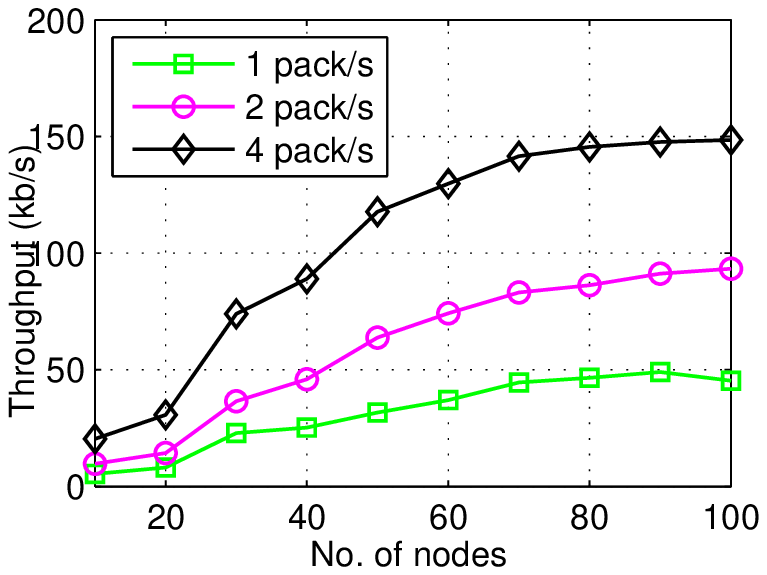}}
  \subfigure[Traffic analysis]{\includegraphics[height=3
  cm,width=4 cm]{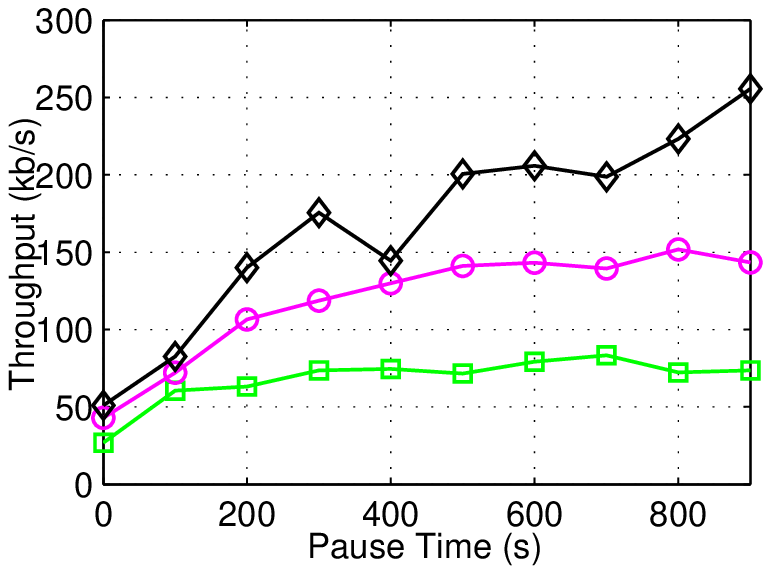}}
  \caption{FSR performance analysis with varying packet rates and scalabilities}
 \end{figure}
\textit{FSR's strange behavior}: \textit{(i) Though it is proactive but its throughput is decreasing with decreasing mobility} because in low mobilities multiple routes are available in \textit{RC}. There is lack of any mechanism to delete expired stale routes in FSR, like DSR, or to determine the freshness of routes when multiple routes are available in route cache, like AODV. \textit{(ii) It is showing least throughput among all protocols}. The reasons include: firstly, for higher traffic rates (large number of packets per second) FSR works well \cite{3-10}. It has been depicted in Fig.5.a. by simulating a scenario with 50 nodes moving at 20m/s speed. It is obvious from Fig that FSR with large number of packets achieves more throughputs. Secondly, FSR is best suited for large scale multi-hop wireless networks, as the \textit{\textit{scope update scheme}} can benefit in reducing the number of routing update packets and achieve high data packet to routing packet ratio. This fact is demonstrated in Fig.5.b. where we have simulated FSR with 20 m/s node speed for varying number of nodes, 10, 20, . . , 100.

\textit{DSR achieves maximum average $throughput_{avg}$} among all six protocols, as, during higher mobility, less RERR messages and RREQ messages are to be sent due to availability of valid routes in \textit{RC}. The promiscuous mode of DSR, as described in Fig.1, step(b), makes able this protocol to handle the high mobility.
\vspace{-0.5cm}
\section{End-to-end Delay (E2ED)}	

	It is the time a packet takes to reach the destination from the source. We have measured it as the mean of Round Trip Time taken by all packets.
\vspace{-0.5cm}
\subsection{E2ED produced by reactive protocols}

As demonstrated in Fig.6, AODV among reactive protocols attains the highest delay. Because \textit{LLR} for link breaks in routes sometimes result in increased path lengths. DYMO produce the lowest $E2ED_{avg}$ among reactive protocols because it only uses the \textit{ERS} for route finding that results less delay; as checking the \textit{RC} in (DSR) and \textit{RT} in (AODV) before route discovery through \textit{ERS} attains a some delay. At higher speeds, DSR suffers the higher AE2ED. The reasons include: for \textit{RD}, it first searches the desired route in \textit{RC} and then starts \textit{RD}, if the search fails. As, DSR does not implement \textit{LLR}, so its AE2ED is less than AODV but during moderate and high mobility at high speed \textit{RC} search fails frequently and results in increased delay.

\begin{figure}[h]
\centering
\includegraphics[
height=6 cm,
width=8.5 cm
]{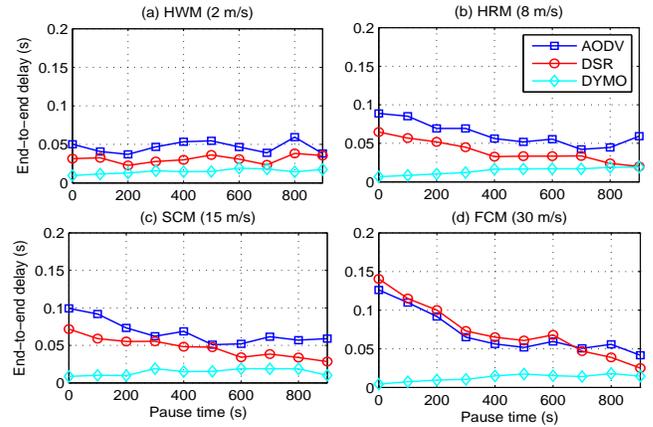}
\caption{End-to-end delay by reactive protocols}
\end{figure}

\vspace{-0.4cm}
\subsection{E2ED produced by proactive protocols}

In all proactive protocols, E2ED value is directly proportional to speed and mobility, as depicted in Fig.7. The proactive protocols have more AE2ED as compared to the reactive ones, as they calculate \textit{RT} before data transmission. DSDV possess the highest E2ED among proactive protocols in moderate and no mobility situations, as well as in all cases its E2ED is higher than OLSR. Because DSDV keeps a data packet until it receives a good route creating delay. Furthermore, advertisements of the routes which are not stabilized yet, is delayed in order to reduce the number of rebroadcasts of possible route entries. 	
	
FSR at higher mobilities, possess the highest AE2ED among proactive protocols. Due to \textit{GF} mechanism when mobility increases, routes to remote destinations become less accurate. However, when a packet approaches its destination, it finds increasingly accurate routing instructions as it enters sectors with a higher refresh rate. At moderate and no mobilities at all speeds, value of end-to-end delay is the same as well as this delay is less than other proactive protocol. It is due to retaining a route entry for each destination, that avoids extra work of "finding" the destination as in on-demand routing.

\begin{figure}[h]
\centering
\includegraphics[
height=6 cm,
width=8.5 cm
]{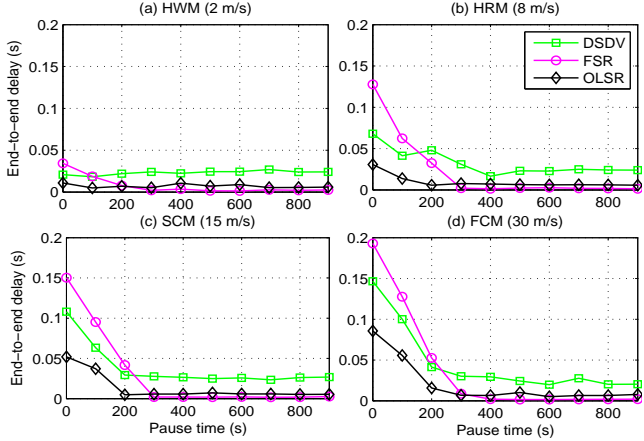}
\caption{End-to-end delay by proactive protocols}
\end{figure}

\vspace{-0.4cm}
\subsection{Interesting facts regarding E2ED}

Generally, reactive protocols cause more delay as compared to the proactive ones. E2ED generated by all 6 protocols is: $AODV>DSR>FSR>DSDV>DYMO>OLSR$, which means that when talking about E2ED, OLSR out performs rest of the five protocols.

\textit{E2ED of DYMO is less not only among the reactive protocols but also from DSDV and FSR} because it neither adapts strategy of \textit{RC} like DSR (step(a), Fig.1) nor \textit{LLR} mechanism like AODV (step(iii), Fig.1). Moreover, DYMO uses \textit{ERS} algorithm which is more efficient for reducing E2ED as compared to \textit{GF} of FSR and waiting for the best route mechanism in DSDV.

	\textit{DSR has the highest E2ED in FCM at moderate and high speed.} Because at high speed, for unreachable destinations, \textit{ERS} algorithm (step(B) of Fig.1.) produces delay to calculate valid routes. As DSR works well in moderate and relatively high rates of mobility, it has to compromise on delay to calculate valid routes.

\textit{AODV suffers from maximum $E2ED_{avg}$.} As, \textit{LLR} mechanism is initiated after link breakage detection. In \textit{RM} phase, step(iii) of Fig.1 is demonstrating that starting of \textit{LLR}, sometimes results in increased path lengths.	

\textit{OLSR achieves the lowest E2ED.} When comparing with proactive protocols, OLSR generates periodic HELLO and Topology Control (TC) messages to check links as well to compute the \textit{MPRs} (\textit{RT calculation phase} step(a) in Fig.2.) to better reduce the delay as compared to periodic exchange of whole table with the neighbors in FSR and periodic and trigger updates in DSDV.

\vspace{-0.5cm}
\section{Normalized Routing Load (NRL)}

NRL is the number of routing packets transmitted by a routing protocol for a single data packet to be delivered successfully at the destination.

\begin{figure}[h]
\centering
\includegraphics[
height=6 cm,
width=8.5 cm
]{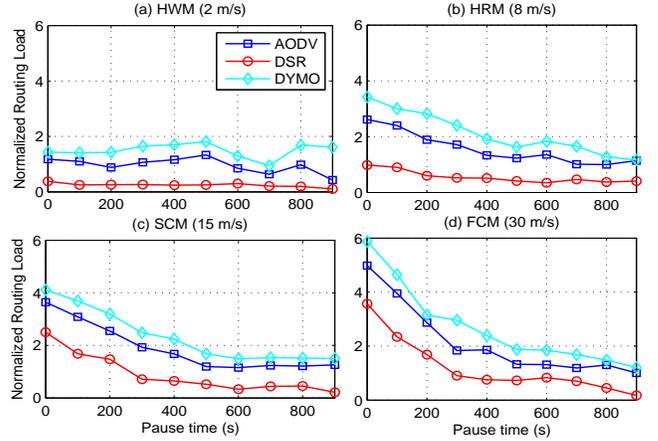}
\caption{Routing overhead by reactive protocols}
\label{fig1}%
\end{figure}

\vspace{-0.4cm}
\subsection{NRL generated by reactive protocols}

Due to the absence of gratuitous RREPs, DYMO produces higher routing overhead than not only reactive protocols but also DSDV and FSR. Whereas, DSR, due to the promiscuous listening mode has the lowest routing load.

\begin{figure}[h]
\centering
\includegraphics[
height=6 cm,
width=8.5 cm
]{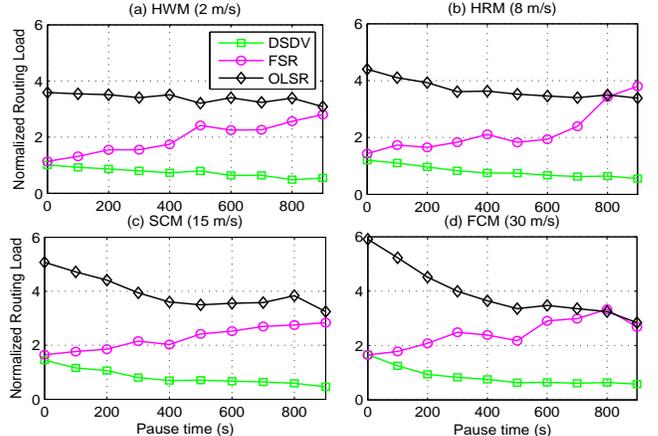}
\caption{Routing overhead by proactive protocols}%
\label{fig1}%
\end{figure}

Although, AODV uses gratuitous RREPs but due to the use of \textit{HELLO messages} like DYMO and \textit{local link repair}, it causes more routing load than DSR. One common noticeable behavior of all reactive protocols is that at high speeds and/or high mobilities, routing overhead is higher as compared to moderate and low mobilities and/or speeds. Because, in response to link breakage, all of the on-demand protocols disseminate RERR message to inform the route request generator about the faulty links and prevent the use of invalid routes. As in high dynamic situations, the link breakage is frequent, so, more RERR messages are generated resulting in high NRL.

\vspace{-0.4cm}
\subsection{Routing overhead produced by proactive protocols}

	Fig. 7 shows that OLSR due to computation of MPRs through TC and HELLO messages results in the highes generation rate of routing packets. The lowest NRL is produced by DSDV, because, \textit{incremental} and \textit{periodic updates} through \textit{NPDUs} reduce the routing overhead. Moreover, FSR has lower routing overhead than OLSR because it prefers \textit{periodic updates} instead of \textit{event driven exchanges} of the topology map which greatly helps in reducing the control message overhead during high mobility rates. Also, in FSR link state packets are not flooded. Instead, nodes maintain a link state table based on the up-to-date information received from neighbor nodes and are periodically exchange it with their local neighbors only (no flooding).

\begin{figure}
  \centering
 \subfigure[Scalability analysis]{\includegraphics[height=3 cm,width=4 cm]{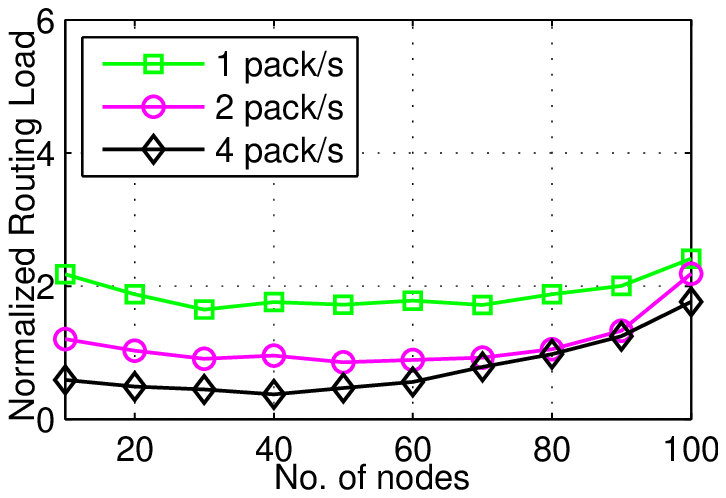}}
  \subfigure[Mobility analysis]{\includegraphics[height=3 cm,width=4 cm]{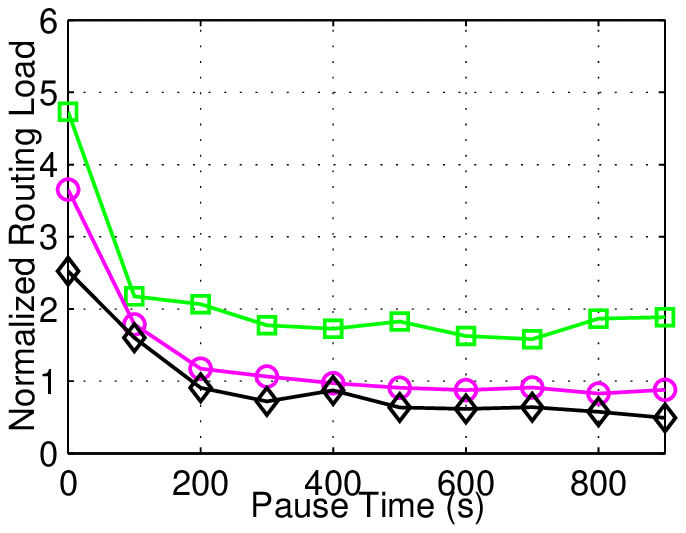}}
  \caption{FSR performance analysis with different packet rates}
 \end{figure}

\vspace{-0.4cm}
\subsection{Interesting facts regarding routing load}

Generally, both classes of protocols; reactive and proactive have to suffer from routing load during higher mobilities and at higher speeds. Following order depicts routing over head of six protocols in which OLSR suffers from the highest number of routing packets: $OLSR>DYMO>FSR>AODV>DSDV>DSR$.

\textit{AODV possesses more NRL than DSR} during all cases of mobility, because an AODV node offers connectivity information by broadcasting local HELLO messages unlike DSR.

\begin{table}[h]
\caption {Performance trad-offs made by routing protocols}
\centering
\tiny\begin{tabular}{|c|c|c|c|c|c|}
\hline
\multirow{2}{*}{\textbf{Protocol}}&\textbf{Modification to}&\multirow{2}{*} {\textbf{Advances achieved}}&
\multirow{2}{*}{\textbf{Price to pay}}\\
        &\textbf{routing technique}&& \\
\hline

  \multirow{2}{*}{\textbf{AODV}}&Sequence number along&Hi thrupt (Fig.3.d, 0s) in&causes delay due to local \\
 &with local link repair&hiest mobility/hiest speed&link repair (Fig.6) \\ \hline

 \multirow{2}{*}{\textbf{DSDV}}&Sequence no. with&hiest thrupt when mobility is&causes delay due to avg.\\
 &avg. settling time&hi and speed is lo (Fig.4.a,0s)&settling time (Fig.7,0s)\\\hline

\multirow{2}{*}{\textbf{DSR}}&Route cache technique&Caches learned routes and &Causes delay when link breaks\\
 & &increase throughput. (Fig.3)&are frequent. (Fig.6.d, $<$600s)\\\hline

 \multirow{3}{*}{\textbf{DYMO}}&Without route cache and &Reduces E2ED in hi mobility &Decreases throughput. (Fig.6.c.d)\\
 &gratuitous route reply&and in high speed. (Fig.3)&and NRL when speed and mobility\\
 &&&is high. (Fig.8.c.d. $<$ 500s)\\\hline

\multirow{4}{*}{\textbf{FSR}}&Multipath routing, &More thruput in hi mobility as &Less throughput and increased\\
 &Fisheye scopes with &comprd to lo moblty.(Fig.4, &E2ED during hi mobility and\\
 &graded frequency & $<$400)  and decrease in NRL.&speed. Fig.8. $>$ 600s and\\
  &mechanism&(Fig.4. $<$ 400s)&Fig.7.b.c.d $<$300s\\\hline

 &&Lo E2ED with more thruput &\\
 \multirow{2}{*}{\textbf{OLSR}}&MPR &(Fig.4, $>$300) in medium or no &Highest NRL, due to MPR's \\
 &calculation&mobility or when speed is lo&computation. (Fig.7.)\\
 &&(Fig.9, $>$300s)&\\\hline

\end{tabular}
\end{table}

\textit{DYMO gives higher NRL value among all reactive protocols} for the reason that though \textit{ERS} algorithm is used to reduce the routing overhead, but AODV and DSR generate \textit{grat. RREP} messages. As demonstrated in step(iii) and in step(c) in \textit{RM phase} of Fig.1, which possibly avoid the second \textit{RD}. These messages are not generated in DYMO causing higher generation rates of routing packets than both AODV and DSR. On the other hand, \textit{RC strategy} further reduces NRL of DSR as compared to AODV.

\textit{FSR's routing overhead is increasing with decrease in mobility.} The reason is that availability of routes in \textit{RC} is inversely proportional to mobility, i.e., in low mobilities more routes are available. There is lack of any mechanism to delete the expired stale routes in FSR or to determine the freshness of routes when multiple routes are available in route cache. These multiple routes not only increase the NRL but also affect throughput. The reasons for strange throughput of FSR are equally valid for routing load. With the same simulation scenarios as carried to justify FSR's strange throughput in Fig.5, we justify the strange NRL of FSR, as shown in Fig.10.

\section{Performance Trade-offs made by protocols}

In this section, referring to routing techniques,upon which the routing protocols are implemented, we compare the performance of the routing protocols they achieve and price they pay. Trade-offs, the routing protocols have to make, are listed in the table.2.

\vspace{-0.4cm}
\section{Conclusion and Future Work}
The massive simulations demonstrate that reactive protocols are superior to proactive ones for mobility constraint. AODV nodes send data packets merely carrying addresses of destination unlike DSR that carry the source routes also. So, DSR has more overhead in bytes than AODV. Whereas, DSR has less overhead in terms of number of packets. AODV broadcasts periodic HELLO messages and sends more control messages than DSR to find and repair the routes by \textit{LLR} technique, so, it produces more routing load than DSR. This can be concluded that AODV and DSR show the best performance during all mobilities and at all speeds. AODV should be chosen where the number of hops is not a problem and the nodes prefer low byte overhead on the packets. For delay sensitive applications, DYMO in reactive protocols and OLSR in proactive protocols are the plausible choices. During all this evaluation, we come to realize that the most important component of a routing protocol is routing link metric, so, future we are interested to propose and implement a new ETX-based routing link metric with AODV and OLSR, as discussed in \cite{3-14}, \cite{3-15}, \cite{3-16}.

\end{document}